%% file: HIIregions_7mm.tex
\documentclass[debug]{rmaa}

\usepackage{paralist}

\usepackage{psfrag,color}

\newcommand{\msun}{M$_{\odot}$}

\newcommand{\accrate}{$\dot{M}$}

\newcommand{\gap}%
{\raisebox{-0.5ex}{$\stackrel{\scriptstyle >}{\scriptstyle \sim}$}}

\title{Exploring the nature of compact radio sources associated to UCHII regions}

\author{
Josep M. Masqu\'e\altaffilmark{1},
Luis F. Rodr\'iguez\altaffilmark{2,3},
Sergio A. Dzib\altaffilmark{4},
S.-N.X. Medina\altaffilmark{4},
Laurent Loinard\altaffilmark{2}, 
Miguel A. Trinidad\altaffilmark{1},
Stan Kurtz\altaffilmark{2}, \&
Carlos A. Rodr\'iguez-Rico\altaffilmark{1} 
} 

\altaffiltext{1}{Departamento de Astronom\'ia, Universidad de Guanajuato}
\altaffiltext{2}{Instituto de Radioastronom\'ia y Astrof\'isica, Universidad Nacional Aut\'onoma de M\'exico}
\altaffiltext{3}{Mesoamerican Center for Theoretical Physics, Universidad Autónoma de Chiapas}
\altaffiltext{4}{Max Planck Institut f\"ur Radioastronomie}

\shorttitle{The nature of compact sources in UCHII regions}
\shortauthor{Masqu\'e et al.}

\fulladdresses{
\item Josep M. Masqué, Miguel A. Trinidad and Carlos A. Rodríguez-Rico: Departamento de Astronom\'ia, Universidad de Guanajuato, Apdo. Postal 144, 36000 Guanajuato, M\'exico

\item Luis F. Rodriguez, Laurent Loinard and Stan Kurtz: Instituto de Radioastronom\'ia y Astrof\'isica, Universidad Nacional Aut\'onoma de M\'exico, Morelia 58089, M\'exico

\item Luis F. Rodriguez: Mesoamerican Center for Theoretical Physics, Universidad
Aut\'onoma de Chiapas, Carretera Emiliano Zapata Km. 4,
Real del Bosque (Ter\'an), 29050 Tuxtla Guti\'errez, Chiapas, M\'exico

\item Sergio A. Dzib and Sac N. Medina: Max Planck Institut f\"ur Radioastronomie, Auf dem H\"ugel 69, D-53121 Bonn, Germany

}

\listofauthors{J. M. Masqué, L. F. Rodríguez, S. A. Dzib, S. N. Medina, L. Loinard, M. A. Trinidad, S. E. Kurtz and C. A. Rodríguez-Rico}
\indexauthor{Masqué, J. M.}
\indexauthor{Rodríguez, L. F.}
\indexauthor{Medina, S. N.}
\indexauthor{Loinard, L.}
\indexauthor{Trinidad, M. A.}
\indexauthor{Kurtz, S. E.}
\indexauthor{Rodríguez-Rico, C. A.}

\abstract{
We present Very Large Array 7 mm continuum observations of four Ultra-Compact (UC)HII regions, observed previously at 1.3 cm, in order to investigate the nature of the compact radio sources associated with these regions. We detected a total of seven compact radio sources, four of them with thermal emission, and two compact radio sources have clear non-thermal emission. The thermal emission is consistent with the presence of an ionized envelope, either static (i.e., trapped in the gravitational radius of an associated massive star) or flowing away (i.e., a photo-evaporative flow). On the other hand, the nature of the non-thermal sources remains unclear and several possibilities are proposed. The possibility that most of these compact radio sources are photo-evaporating objects and the remaining ones more-evolved objects is consistent with previous studies on UCHII regions. 
}

\resumen{Presentamos observaciones del continuo a 7 mm usando el Very Large Array de cuatro regiones HII Ultra-Compactas (UCHII), observadas previamente a 1.3 cm, con el fin de investigar la naturaleza de las fuentes de radio compactas asociadas con estas regiones. Detectamos un total de siete radio fuentes compactas, cuatro de ellas con emisión térmica, y dos radio fuentes compactas tienen claramente emisión no térmica. La emisión térmica es consistente con la presencia de una envolvente ionizada estática (i.e., atrapada en el pozo de potencial de la estrella masiva asociada a la región) o fluyendo (i.e., un flujo foto evaporándose). Por otro lado, la naturaleza de las fuentes no térmicas no es clara y se proponen algunas posibilidades. La posibilidad de que la mayoría de estas fuentes de radio compactas sean objetos foto evaporándose y las fuentes restantes objetos más evolucionados es consistente con estudios previos sobre regiones UCHII.
}

\addkeyword{stars: formation}
\addkeyword{HII~regions}

\begin{document}

\maketitle

\section{Introduction}

The advent of a new generation of upgraded instruments (e.g.\ the Karl G. Jansky Very Large Array) enables radio astronomers to detect and systematically study the weakest and most compact sources emitting at radio wavelengths in star forming regions. As an outcome, extremely rich populations of cm emitting sources, most of them a few hundreds of AU in size (when resolved), are unveiled in these regions \citep[e.g.,][]{forbrich2016,medina2018}. The nature of these Compact Radio Sources (CRS) is not unique as shown by the variety of spectral indices that they exhibit, among other properties. Furthermore, special attention has been recently paid to the particular case of CRS associated with UCHII regions (i.e., very near to the massive star). Good examples can be found in W3(OH) \citep{kawamura1998, dzib2013b} and NGC6334A \citep{carral2002, rodriguez2014}. The detection of these sources associated with the probably densest  part of the cloud is challenging because they are usually weak compared to the extended free-free emission of the UCHII region where they are possibly embedded. In a recent systematic study \citep[][hereafter MRT2017]{masque2017}, 12 UCHII regions were observed to obtain an unbiased census of CRS associated with them. They reported 13 CRS, showing that the above-mentioned association is common. The significant number of CRS found allowed MRT2017 to classify them into two main types depending on the position of the compact sources with respect to the peak emission of the UCHII region. The CRS that appear projected close to the peak (i.e., possibly embedded in the dense ionized gas of the region) were called Type I sources, whereas the CRS scattered around the UCHII region were called Type II sources. The authors argue that Type I sources probably correspond to photo-evaporating objects with high mass depletion rates as a consequence of the harsh irradiation that they are suffering from the nearby massive star. The large amount of expanding plasma emanating from these objects, either associated or not with the exciting star, maintains large emission measures in a compact volume around it,  and an UCHII region is observed \citep[lasting $\sim10^5$ yr according to][]{hollenbach1994}.  On the other hand, Type II sources could be photo-evaporating objects with lower depletion rates because the massive star is not so nearby. Their low mass depletion rates would prevent producing an observable UCHII region around them and the CRS appears 'naked'. As they are probably long-lived, Type II objects are expected to be on average more evolved than Type I objects, and some of them could be pre-main sequence stars.


In this paper we present a follow-up study of MRT2017 to explore further the nature of the CRS. The interpretation of the nature of CRS given in MRT2017 was tentative, but sufficient to provide a possible schematic evolution and elucidate the role of CRS  in star forming regions. We will constrain observational properties of the CRS, particularly the spectral index, supported by new Q-band observations of 4 UCHII regions containing both types of CRS. The results presented here are important to probe the scheme proposed in MRT2107 and shed light on the evolution of young objects embedded in giant clouds. The 7 mm observations are described in Sect. 2. In Sect. 3 we give our observational results and their discussion is presented in Sect. 4. Finally, in Sect. 5 we list our conclusions.


\section{Observations and Data Reduction}

\subsection{Source Selection}

We observed the UCHII regions G28.29-0.36, G35.20-1.74, G60.88-0.13 and G61.48+0.09A. All these regions are excited by late O or early B type stars and are part of the original selection of MRT2017. These authors found that G28.29-0.36 and G35.20-1.74 contain CRS of Type I, whereas G60.88-0.13 and G61.48+0.09A have Type II CRS. Our UCHII regions are located at fairly high declinations since at 7 mm the quality of the observations strongly depends on source elevation. At the same time, we excluded the most compact UCHII regions because the CRS can be difficult to isolate in their corresponding maps.

G28.29-0.36, located at 3.1 kpc, is associated with IRAS 18416-0420. The maps of \citet{kurtz1994} show that this UCHII region is composed of two peaks aligned north-south, with the CRS associated with the northern one. The G35.20-1.74 UCHII region is part of the W48 complex \citep{zeilik1978} located at 3.3 kpc \citep{zhang2009}. The cm emission at subarcsec scales shows a cometary shape with the tip oriented to the NE \citep{wood1989, kurtz1994}. The CRS is found less than $1''$ away from the peak of the UCHII region and very close to its geometric center. The G60.88-0.13 region, or Sh 87,  is located at 2.1 kpc. Several IR sources were found in the field \citep{campbell1989}, suggesting that a cluster of YSO surrounding the UCHII region is present. Within uncertainties, the CRS is coincident with one of these IR sources. Finally, G61.48+0.09A is embedded in the emission nebula Sh 2-88B, for which we adopted a distance of 2.0 kpc \citep{crampton1978}.  This UCHII region presents compact northern emission and some southern extended emission. We found two CRSs embedded in the southern part, which is filtered out by our observations, and we consider these CRS detached from the densest part of the UCHII region. One of them is resolved with an arc-shaped morphology.


\subsection{7 mm Continuum Observations and Mapping Procedures}

On 2016 May 22 and 24, we observed four UCHII regions of the sample of 12 regions observed at 1.3 cm by MRT2017 (program VLA/16A-003). The observations were performed in the 7 mm band with the VLA in the B configuration. We tuned the frequency range of 40.0-47.9 GHz arranged in 64 spectral windows of 64 channels each. Each channel had a width of 2 MHz. We used the 3 bit samplers with full polarization and 3 seconds as the integration time. The flux and bandpass calibrator was 3C286. Pointing corrections were applied for telescope slews larger than 10 degrees on the sky. We employed integrations of 1 to 1.5 minutes on-source preceded and followed by integrations of 25 to 40 sec on the gain calibrator to obtain a total on-source time of about 6 minutes. This sequence permits calibration solutions in intervals below the typical time scale of atmospheric phase fluctuations (which can be highly variable at the high frequency bands of the VLA) due to turbulence in the troposphere. The observational parameters are listed in Table \ref{observations}.

The data were calibrated with the Common Astronomical Software Applications (CASA) package through the pipeline provided by NRAO. The CLEAN task of CASA with the \emph{nterms} parameter set to 2 was used to construct the maps. We obtained a first set of maps setting the robust parameter to 0 to search for all the emission components. In order to remove extended emission from the maps, we re-imaged the fields using only uv-distances larger than 515 k$\lambda$, equivalent to mapping only structures smaller than 0\rlap{$''$}.5 in size, and with uniform weighting. This cut was chosen to employ the same uv-range as that used in the synthesis of the 1.3 cm maps of MRT2017. This is a requirement for a proper comparison between the maps at both wavelengths in the forthcoming analysis. The parameters of the 7 mm maps are shown in Table \ref{maps}.

\section{Results}
  
\begin{figure}[thbp]
\centering
\resizebox{1.1\textwidth}{!}{\includegraphics[angle=0]{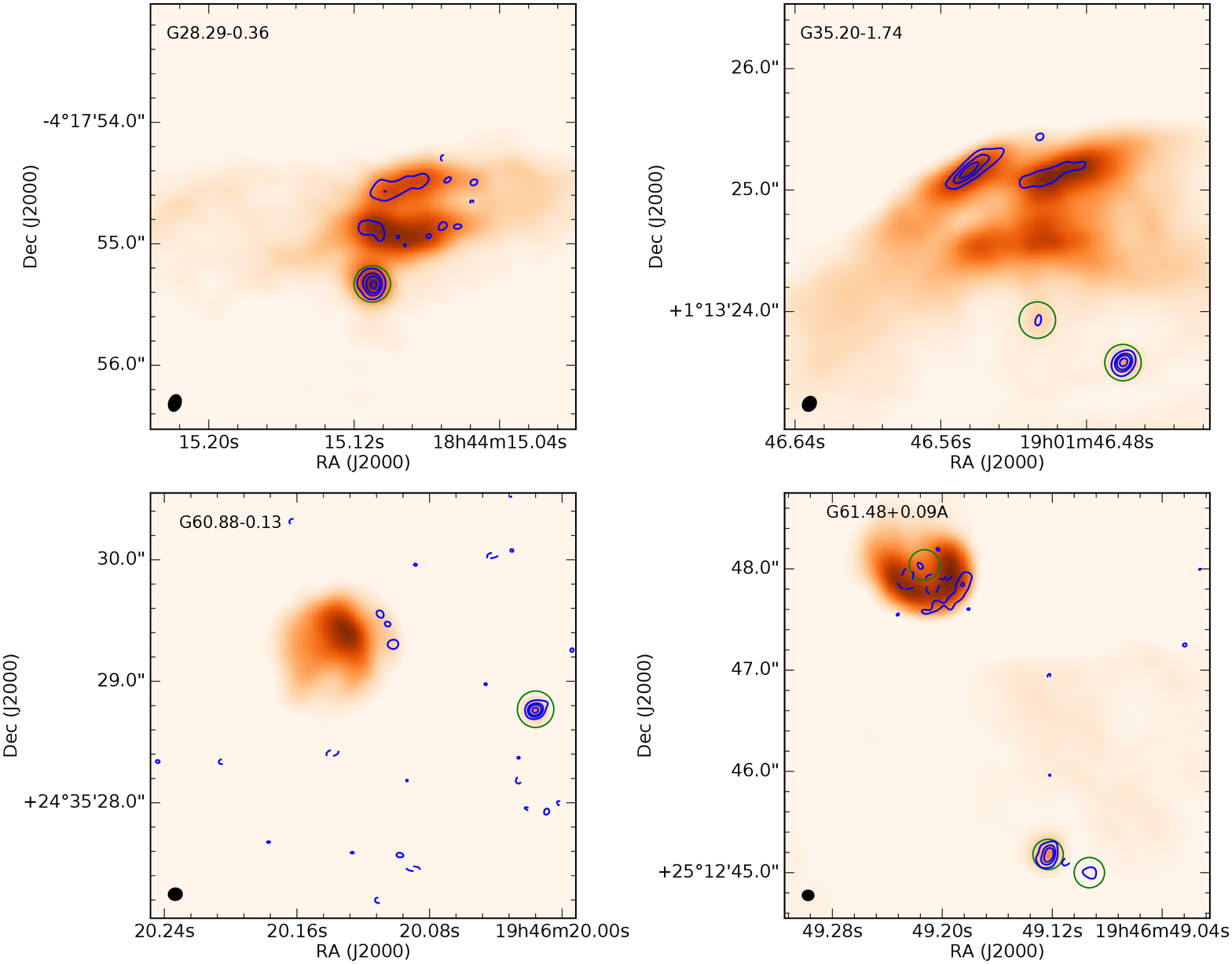}}
\caption{Continuum 7 mm emission maps of the observed regions listed in Table \ref{observations} constructed with all the visibilities (colour scale) with the maps constructed with baselines longer than 515 k$\lambda$ superimposed (contours). For the contours, levels are -3, 3, 7, 10, and 15 times the $rms$ noise level shown in Column 6 of Table \ref{maps}. The colour-scale maps are smoothed to an angular resolution of 0.2$''$ to improve the appearance. The green circles indicate the location of the CRS reported in Table \ref{sources_flux}. The synthesized beam of the maps without short spacings is shown in the bottom right corner. 
\label{general_maps}}
\end{figure}

\begin{figure}[thbp]
\centering
\resizebox{0.9\textwidth}{!}{\includegraphics[angle=0]{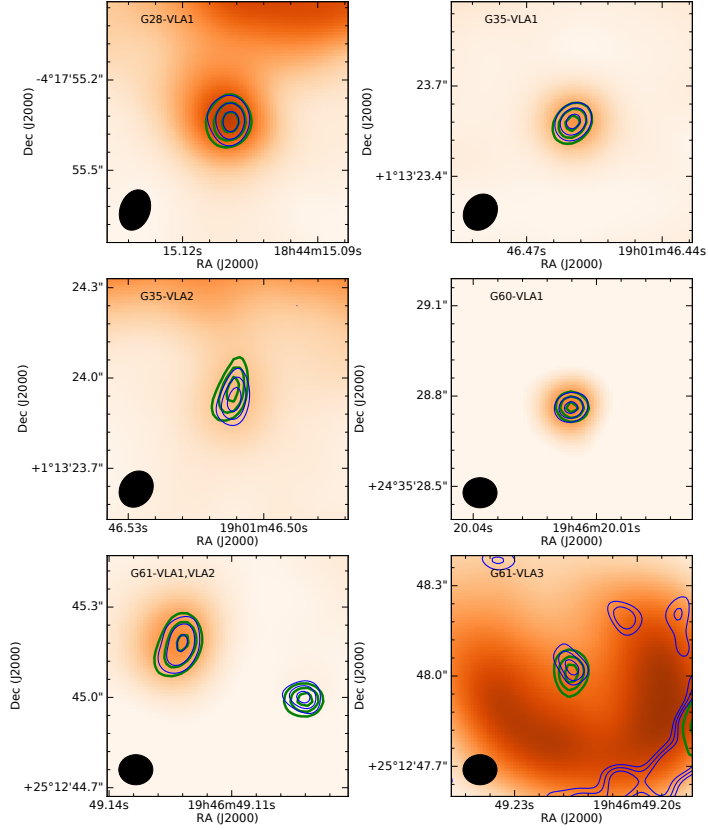}}
\caption{Zooms on the CRS indicated by green circles in Figure \ref{general_maps}. The color scale represents the 7 mm maps constructed with all the visibilities; the blue contours represent the levels corresponding  to 50\%, 70\% and 90\% of the brightness peak of the CRS and the green contours represent the same but for 1.3 cm continuum emission. The maps represented in color scale are the same as those shown in Figure \ref{general_maps} (i.e. smoothed to an angular resolution of 0.2$''$) but with a modified range. Both maps represented in contours have been convolved to the same synthesized beam which is shown in the bottom left corner.   
\label{detail_maps}}
\end{figure}

In Figure \ref{general_maps} we show our VLA 7 mm maps. The color scale corresponds to the maps obtained including all the visibilities while the contours show only the longer baseline data. The G28.29-0.36 and G35.20-1.74 regions are more extended than the nominal largest angular scale structure visible to the array ($4''$ observing at the 7 mm band in the B configuration), hence we only recover the brightest emission of these regions. On the other hand, the maps without extended structures shown in contours appear cleaner and the compact structures of the regions are clearly seen. In these maps, all the CRS reported in MRT2017 for these four UCHII regions are detected at 7 mm, confirming that they are real structures and not artifacts of the maps. Using the nomenclature of MRT2017, their are G28-VLA1, G35-VLA1, G60-VLA1, G61-VLA1 and G61-VLA2; they are shown in Figure \ref{detail_maps}.

We found two additional CRS not reported in MRT2017 (we name them G35-VLA2 and G61-VLA3 and they are also shown in Figure \ref{detail_maps}). To confirm these new detections, we inspected the 1.3 cm maps of the corresponding UCHII regions and found that these new sources were marginally detected in the MRT2017 maps. They both share the characteristic of being centered on the UCHII region and they could be associated with the exciting source. The detection of additional CRSs agrees with the scenario that these sources are common in the surroundings of the UCHII regions, most of them being extremely weak and hard to detect.

As a comparison, the contours of the 1.3 cm continuum maps convolved to the larger 7 mm beam are also displayed in Figure \ref{detail_maps}. As a general trend, the compact sources are unresolved or marginally resolved at both wavelengths and, thus, they appear very similar in the maps. Only G35-VLA2 shows an offset of $\sim0\rlap{$''$}.07$ between the two frequencies. This displacement is smaller than our estimated positional error and so it is unlikely to be real.

Table \ref{sources_flux} shows the measured flux densities at 7 mm (and 1.3 cm using the MRT2017 data) and spectral indices for the seven CRS. Flux densities and peak intensities were derived by fitting two dimensional Gaussians to each CRS in the same manner for both bands
 (with the 1.3 cm maps convolved to the 7 mm beam size). From these parameters we derived spectral indices for the integrated flux densities of the CRSs ($\alpha_S$) and for their peak intensities ($\alpha_I$). If the source is resolved, these two indices might differ if inhomogeneities in the source structure are present. Table \ref{deconv} shows the deconvolved sizes (angular and linear) of the three CRSs that were resolved, at least marginally, in the fit. Despite the large errors, an assessment of the true physical size of the source can be obtained. Moreover, in Table \ref{deconv} we show physical parameters derived from the flux density and size of the CRS, namely, brightness temperature, optical depth and emission measure. As can be seen in Table \ref{deconv}, G35-VLA1 appears to be smaller and denser than the other sources. These physical parameters are consistent with those derived in MRT2017. Thus, the 1.3 cm and 7 mm continuum emission arises from the same region, at least in the resolved sources.

  
\section{Analysis of selected sources}

In this section we analyze the nature of the 7 CRS detected in our 7 mm observations. A close examination of the properties of these CRSs listed in Tables \ref{sources_flux} and \ref{deconv} shows that their properties are varied suggesting different physical nature for these sources. Thus, in the following we analyze them individually:

\emph{G28-VLA1}: The possibility that this source is a very small HCHII region was discarded by MRT2017 because they show that G28-VLA1 is unlikely to host the massive star exciting the region. The remaining possibilities are externally photo-evaporated disk/clump winds and thermal radio jets associated with young low-mass objects. Despite the different launching mechanism and structure of the outflowing gas, their observational signposts are often very similar. Unfortunately, our angular resolution yields a deconvolved size too uncertain for an accurate morphological inspection (either at 1.3 cm or 7 mm). Furthermore, our observing frequency is well above the turn-over frequency ($\sim$10 GHz for G28-VLA1) as indicated by the low values of optical depth ($\sim4 \times 10^{-2}$). This results in spectral indices  $\alpha_I$ ($\sim0$) and $\alpha_S$ ($\sim-0.2$) consistent with optically thin thermal emission of ionized gas. Flat spectral indices are expected for jets and disk winds when observed at frequencies high enough that essentially the whole object becomes optically thin  \citep{reynolds1986, hollenbach1994, anglada2018}. 

On the other hand, the emission measure of G28-VLA1 ($3 \times 10^8$ cm$^{-2}$ pc) is in excellent agreement with the typical values obtained by \citet{hollenbach1994} for their photo-evaporating disk wind models but is significantly higher than that expected for a jet \citep[e.g, see][]{rodriguez1990, curiel1993}. Furthermore, jets usually show additional structures such as knots resulting from previous ejecta \citep[e.g.][]{marti1993}, which are not seen in our maps. Therefore, we favor the possibility that the optically thin emission of G28-VLA1 is produced by a photo-evaporated disk/clump wind.

\emph{G35-VLA1}: This CRS has the largest derived $EM$ ($\sim10^9$ cm$^{-6}$ pc) and a relatively compact size ($<100$ AU) compared to other resolved CRSs (G28-VLA1 and G61-VLA1). Based upon size and density, MRT2017 proposed that G35-VLA1 is associated with the exciting star of the UCHII region, as its size is smaller than twice the gravitational radius. The large emission measure of G35-VLA1 discards the jet possibility for this CRS. The turn over frequency for an emission measure of $2.2 \times 10^9$ cm$^{-6}$ pc is $\sim$25 GHz. At our observing frequency of $45$ GHz the gas should be completely transparent. This seems consistent with the flat values for the spectral indexes. However, within the uncertainties, we neither can discard a slight step on these indices and the optical depth ($\sim$0.3) is somewhat large to be consider optically thin emission. 

By comparing $\alpha_S$ and $\alpha_I$  of Table \ref{sources_flux}, we see that $\alpha_S  <  \alpha_I$. The same result was found for some of the radio sources of Cepheus A by \citet{garay1996}. These authors conclude that a larger spectral index of the peak emission suggests the presence of an optically thick compact region at the center of the source. Considering G35-VLA1 a trapped HCHII region, its optically thick part must be $<$ 70 AU in radius \citep[see Table 5 of][where an O6 star is assumed, more massive than the star exciting G35.20-1.74]{keto2003}. On the other hand, following the prescription of \citet{hollenbach1994} for photo-evaporating disk winds (see their equation 7.1) this radius goes between 36-168 AU observing at 7 mm and for mass loss rates typical for Type I objects ($10^{-5} - 10^{-6}$ \msun\ yr$^{-1}$). These values are consistent with the deconvolved size of G35-VLA1 of Table \ref{deconv}. Because this size is well below our angular resolution only a slight rise in $\alpha_I$ is observed. Therefore, the observational properties of G35-VLA1 suggest a density gradient in the CRS that could be the result of outflowing ionized material or a trapped HCHII region.

\emph{G35-VLA2}: This CRS lies near the geometrical center of the G35.20-1.74 UCHII region and is clearly a Type I object according to MRT2017. The detection of G35-VLA2 is too marginal to reach any strong conclusion about its nature; additional observational constraints are required.

 \emph{G60-VLA1}: MRT2017 propose that this source is composed of ionized gas trapped in the potential well of a B-type star forming a small HCHII region. This is consistent with the lower limit of EM of $7 \times 10^7$ cm$^{-6}$ pc for this CRS. Moreover, our derived spectral index is $\sim1.7$, which indicates optically thick gas, usually found at centimeter wavelengths in HCHII regions.

\emph{G61-VLA1}: This CRS is the best-resolved of our sample and shows hints of an arc-like shape. The tip of the arc points towards G61-VLA2 suggesting that the latter is responsible for photo-evaporating the former. Both $\alpha_S$ and $\alpha_I$ for G61-VLA1 are clearly negative, showing unambiguously that the emission is mainly non-thermal. Other cases of negative spectral indices around proplyds have been previously reported \citep{mucke2002, masque2014}, and are attributed to the presence of a relativistic electron population in a magnetized medium. Regardless of the mechanism that produces high-energy electrons, an estimation of the magnetic field can be obtained by assuming equipartition of energy between the magnetic field and relativistic particles. Following \citet{pacholczyk1970}, the magnetic field estimated under these conditions is given by:



\begin{equation}
B [\rm Gauss] = \left [ 4.5 ( 1 + \chi ) c_{12} L \right ]^{2/7} R^{-6/7}  
\end{equation}

\noindent where $L$ and $R$ are the radio luminosity and radius of the source (assuming spherical symmetry) given in $cgs$ units, and $\chi$ is the energy ratio of the heavy particles to electrons. $L$ is obtained by integrating the area below the spectra derived from the 1.3 cm and 7 mm fluxes and extrapolated to the limits of 0.01 to 100 GHz \citep[e.g., ][]{miley1980}. Thus, this parameter implicitly includes the dependence on the spectral index $\alpha$. The total luminosity in the radio domain is $1.6 \times 10^{30}$ erg s$^{-1}$.  
The parameter $c_{12}$ is a function of the spectral index and frequency with tabulated values given by \citet{pacholczyk1970}. In our particular case its value is $2\times10^7\,{\rm Gauss}^{3/2}{\rm s}$. For $R$ we calculated an equivalent source size $\theta_\mathrm{eq}=(\theta_\mathrm{M}\theta_\mathrm{m})^{1/2}$ (see Column 2 of Table \ref{deconv}) which corresponds to a linear size of $R= 1.4 \times 10^{15}$ cm. The value of $\chi$ is very uncertain and varies from 1 to 2000 depending on the mixture of relativistic particles considered (e.g. electrons, positrons and/or protons). However, the exponent 2/7 makes the magnetic field weakly dependent on $\chi$ and we do not expect changes beyond an order of magnitude. For this parameter, we adopted 40 as measured by \citet{simpson1983} near the Earth.

We obtained a value of $B \sim 25$ mG that, despite being a rough number (it has about 50\% uncertainty), it is well above other values found in a variety of radio sources showing thermal emission such as Cepheus A \citep[$\sim0.3$ mG,][]{garay1996} or in proplyds like those found in the NGC 3603 region \citep[$\sim 1$ mG,][]{mucke2002}. The high value of $B$ depends on the compact size of G61-VLA1 and suggests that the magnetic field increased its strength as a consequence of a prior contraction  (either gravitational or driven by external pressure) suffered by these sources \citep{mouschovias1976}. Consistently, the proplyds discussed in \citet{mucke2002} are much larger than G61-VLA1 ($10^4$ vs. 200 AU). The tendency of $\alpha_S  <  \alpha_I$ can be interpreted as an increase of electron density in the central part of the source, similar to the G5.59-VLA1 proplyd  \citep{masque2014}: given a mixture of thermal and non-thermal emission, the former will tend to dominate in the densest part of the source because it depends on n$_e$$^2$, where n$_e$ is the electron density, contrary to non-thermal emission that depends on other parameters like magnetic field or shock energetics.         
      
Alternatively, a wind collision region (WCR) of two massive stars, where the most massive member is G61-VLA2, could explain the non-thermal nature of the emission and the morphology of G61-VLA1. Examples of WCR between two massive stars have been reported previously, such as in Cyg OB \#5 \citep{contreras1996,contreras1997,dzib2013a}. This scenario is favored by our value for the magnetic field that matches well with that found in other WCR \citep[20 mG in HD 93129A,][]{delpalacio2016}. 
The only caveat to this scenario is the lack of a point IR counterpart indicating the presence of a massive ionizing star in G61-VLA2, unless its IR emission is obscured by a small disk located edge on.  A possible candidate is a 2MASS point source located $10''$ (20,000 AU at 2 kpc) to the west. With the present data, we cannot discriminate between a magnetized neutral object (i.e., a clump or protostellar disk) that is being photo-evaporated or a WCR between massive stars present in the region to explain the nature of G61-VLA1.  




\emph{G61-VLA2}: The spectral index of G61-VLA2 is similar to that of collimated winds (i.e., jets) or isotropic winds with a density gradient\citep [0.6, ][]{panagia1975}. The lower limit of EM for this CRS ($4 \times 10^7$ cm$^{-6}$ pc) rules out the protostellar jet possibility as they have lower expected values \citep{rodriguez1990, curiel1993}. On the other hand, an expanding ionized disk wind with a radial density gradient is an excellent match (in terms of spectral index) with the \citet{hollenbach1994} model. Therefore, G61-VLA2 could be a photo-evaporating object, possibly a disk around a YSO. However, instead of having a photo-evaporated flow with a resolved volume (sizes $> 200$ AU) such as other CRS, the volume of the ionized flow of G61-VLA2 is so small that its optically thick region fills a significant fraction of the total volume. The nearby cometary-shaped G61-VLA1 supports the existence of a stellar/disk wind.

\emph{G61-VLA3}: Despite its large uncertainty, the spectral index of G61-VLA3 is clearly negative ($\sim -1$). This suggests gyro-synchrotron radiation produced in the magnetically-active corona of a young solar type star.  This emission mechanism is commonly observed in star forming regions containing low-mass YSO \citep[e.g.,][]{dzib2013a}.  Since they can show dramatic flux variations from days to hours, we obtained fluxes separately for the two observing days and compared their values to search for rapid flux variability.  We observed differences of 40\%, which is slightly above the relative errors of our flux determination and consistent with a non-thermal nature of G61-VLA3. Nevertheless, future flux monitoring of this source is needed to assess flux variability. Also, we mapped the Stokes $V$ parameter to look for circular polarization at 1.3 cm and 7 mm. The weak brightness of this CRS hinders a clear detection of polarized emission, but we obtained an upper limit of 10\%, from which we cannot discard the presence of circularly polarized emission and the possibility that G61-VLA2 is a magnetic star. 


\section{Discussion and conclusions}

We detected seven CRS located in four UCHII regions previously studied by MRT2017. The analysis of their properties yields that four of them have thermal emission and two have a non-thermal nature. One remaining source (G35-VLA2) is detected too marginally to elucidate its emission nature.

The four thermal sources have free-free emission arising from ionized gas in the CRS. According to the MRT2017 scenario, G28-VLA1 and G35-VLA1 are Type I sources and, hence, the ionized gas corresponds to a photo-evaporated flowing material with high mass depletion rates (\accrate$ >10^{-6}$ \msun\ yr$^{-1}$).  
On the other hand, G60-VLA1 and G61-VLA2 are Type II sources and have significantly lower or even no depletion rates. This scenario is in excellent agreement with our interpretation of the nature of G28-VLA1 as it appears well-centered on G28.29-0.36 and, at least partially, is feeding the UCHII region with its photo-evaporated outflowing material. The scenario is consistent with the isolated locations of CRSs G60-VLA1 and G61-VLA2, as they are unable to produce a sufficient extended ionized region around them. It is worth noting the possibility that G60-VLA1 is a trapped HII region is also consistent with the scenario discussed by \citet{keto2002,keto2003}, where instead of outflowing gas, the accretion proceeds through the HII region and the star is still growing in mass. 

There is some ambiguity in the interpretation on the nature of G35-VLA1 between being a trapped HII region or a photo-evaporated wind. Depending on each possibility, the material within the UCHII region could be expanding or infalling from/onto the CRS. Interestingly, these scenarios are not mutually exclusive if rotation is included \citep{keto2006,keto2007}. In the resulting configuration, the gas is outflowing except for a given range of equatorial angles with respect to the rotation axis, where the gas is infalling. This possibility implies bipolar morphologies that are not observed for G35-VLA1 because its compact size prevents us from determining its morphology with these observations. Additional observations with higher angular resolution (i.e.,  the A configuration) to explore the kinematics of the G35.20 UCHII region (i.e., through recombination lines) are crucial to shed light on this last possibility.

The non-thermal CRSs possibly have natures different than the photo-evaporating objects. G61-VLA1 is a Type II source and, hence,  consistent with the MRT2017 scenario. On the other hand, the  
location of G61-VLA3 centered on G61.48+0.09 and the nature of its radio emission mechanism are puzzling: it is difficult to explain the presence of a low mass YSO embedded in an UCHII region (i.e., Type I source). A possible scenario can be that G61-VLA3 is unrelated to the dense gas of the UCHII region but projected in its direction on the sky. Such an arrangement is unlikely but not impossible \citep[e.g.][]{trejo2010, dzib2016}. Future investigations to constrain the nature of these CRS are required to conclusively confirm or reject the scenario proposed in MRT2017.

\bibliography{HIIregions_7mm}
\newpage

\input{observations_7mm.tex}

\input{maps_7mm.tex}

\input{sources_flux.tex}
\input{sources_deconv.tex}

\end{document}

%% file: observations_7mm.tex
\begin{table}[ht]
\footnotesize
\begin{center}
\caption{Observed UCHII regions and summary of the VLA observations}\label{observations}
\begin{tabular}{p{1.9cm}ccccccc}
\hline
\hline
UCHII & Distance$^{\mathrm{a}}$& \multicolumn{2}{c}{Pointing Center}   & Gain &  Positional Accuracy$^\mathrm{b}$ & $S_\mathrm{\nu}^\mathrm{cal,c}$& Spectral \\
 name   & (kpc)& $\alpha$~(J2000) & $\delta$~(J2000) &  Calibrator & (mas) &  (Jy) &  Index$^{\mathrm{d}}$  \\
\hline
G28.29-0.36 &  3.1 & $18^\mathrm{h}44^\mathrm{m}15^\mathrm{s}097$  & $-04^{\circ}17'55\rlap{$''$}.29$ &  J1851-0035 &150 & 0.68  &  -0.54 \\
G35.20-1.74 &  3.3 &$19^\mathrm{h}01^\mathrm{m}46^\mathrm{s}490$  & $01^{\circ}13'24\rlap{$''$}.65$ &  J1851-0035& 150 & 0.68  & -0.54\\
G60.88+0.13 &  2.1 &$19^\mathrm{h}46^\mathrm{m}20^\mathrm{s}130$ & $24^{\circ}35'29\rlap{$''$}.39$ & J1931+2243 &  2&0.36 & -0.41  \\
G61.48+0.09 & 2.0  &$19^\mathrm{h}46^\mathrm{m}49^\mathrm{s}202$ & $25^{\circ}12'48\rlap{$''$}.05$ &  J1931+2243& 2 &0.36& -0.41\\
\hline
\end{tabular}
\end{center}

$^{\mathrm{a}}${See references in the text.} \\
$^{\mathrm{b}}${Obtained from the NRAO database} \\
$^{\mathrm{c}}${Bootstrapped flux density of gain calibrator at 41.128 GHz.}\\
$^{\mathrm{d}}${Spectral index of gain calibrator derived from the outcome of CASA tasks.}
\end{table}

%% file: maps_7mm.tex
\begin{table}[ht]
\footnotesize
\begin{center}
\caption{Parameters of the VLA maps}\label{maps}  
\begin{tabular}{lccccccc}
\hline
\hline
& \multicolumn{2}{c}{Maps with all the visibilities$^{\mathrm{a}}$} &  & \multicolumn{2}{c}{Maps without short spacings$^{\mathrm{b}}$} & \\\cline{2-3}\cline{5-6}
& Beamsize  & map $rms$ noise   & &Beamsize& map $rms$ noise &   \\
UCHII name   &  ($'' \times ''$; $^\circ$) & ($\mu$Jy bm$^{-1}$) & &  ($'' \times ''$; $^\circ$) & ($\mu$Jy bm$^{-1}$) & $N_\mathrm{sources}^{\mathrm{c}}$\\
\hline
G28.29-0.36  &  $0.160 \times 0.119$; -19.2 & 120 && $0.138 \times 0.099$; -19.7& 70 & 1 \\
G35.20-1.74 &  $0.148 \times 0.124$; -33.2 & 300  && $0.127 \times 0.106$; -35.2& 90 & 2\\
G60.88+0.13 & $0.130 \times 0.116$; -84.3 & 45  && $0.113 \times 0.100$; 89.4& 60 & 1\\
G61.48+0.09 &  $0.130 \times 0.117$; -89.4 & 80 && $0.113 \times 0.100$; -89.7& 50 & 3 \\
\hline
\end{tabular}
\end{center}
$^{\mathrm{a}}${Maps weighted with robust 0. } \\ 
$^{\mathrm{b}}${Maps constructed with uniform weighting. We removed structures larger than 0\rlap{$''$}.5 (see text).} \\ 
$^{\mathrm{c}}${Number of CRS associated with the region. We included the newly discovered sources.} \\ 
\end{table}

%% file: sources_flux.tex
\begin{table}[ht]
\footnotesize
\begin{center}
\caption{Observed fluxes and spectral indices for the CRSs} \label{sources_flux}
\begin{tabular}{p{1.7cm}cccccccc}
\hline
\hline
& \multicolumn{2}{c}{Coordinates$^a$}    &    $S_\mathrm{7mm}$  & $I_\mathrm{7mm}^{Peak}$ & $S_\mathrm{1.3cm}$$^b$  & $I_\mathrm{1.3cm}^{Peak}$$^b$ && \\
Source   &  $\alpha$~(J2000) & $\delta$~(J2000)   &  (mJy) & (mJy bm$^{-1}$) &   (mJy) & (mJy bm$^{-1}$)& $\alpha_S$$^c$ & $\alpha_I$$^d$ \\
\hline
G28-VLA1 & 18:44:15.110 & -4.17.55.33 & $ 3.0  \pm 0.2$ & $ 1.79 \pm 0.08$ &   $3.41 \pm 0.22$ & $1.79 \pm 0.08$           & $-0.18 \pm 0.13$ & $0.00 \pm 0.09$    \\
G35-VLA1 & 19:01:46.460 & 01.13.23.58 & $ 5.7  \pm 0.2$ & $ 4.8 \pm 0.1$ &   $5.88 \pm 0.23$ & $4.53 \pm 0.11$        & $-0.05 \pm 0.07$ & $0.09 \pm 0.05$  \\
G35-VLA2 & 19:01:46.507 & 01.13.23.93 & $1.0 \pm 0.4$ & $ 1.09 \pm 0.21$  &   $1.0 \pm 0.3$ & $0.89 \pm 0.16$         & $-0.1 \pm 0.7$ & $0.3 \pm 0.4$  \\
G60-VLA1 & 19:46:20.016 & 24.35.28.77 & $ 1.01  \pm 0.07$ & $ 1.04 \pm 0.04$ &   $0.31 \pm 0.02$ & $0.34 \pm 0.01$        & $1.69 \pm 0.13$ & $1.62 \pm 0.07$ \\
G61-VLA1 & 19:46:49.123 & 25.12.45.18 & $ 2.4  \pm 0.3$ & $ 1.1 \pm 0.1$  &   $3.5 \pm 0.3$ & $1.39 \pm 0.09$      & $-0.56 \pm 0.21$ & $-0.28 \pm 0.15$  \\
G61-VLA2 & 19:46:49.093 & 25.12.45.00 & $ 0.6  \pm 0.1$ & $ 0.54 \pm 0.06$ &   $0.40 \pm 0.06$ & $0.35 \pm 0.03$       & $0.5 \pm 0.3$ & $0.61 \pm 0.20$ \\
G61-VLA3 & 19:46:49.216 & 25.12.48.03 & $0.28  \pm 0.07$  & $0.28 \pm 0.07 $ &   $0.73 \pm 0.16$ & $0.56 \pm 0.08$      & $-1.4 \pm 0.5$ & $-1.0 \pm 0.4$  \\
\hline
\end{tabular}
\end{center}
$^a${Typical positional uncertainties are $\sim150$ mas for the CRSs associated with G28.29-0.36 and G35.20-1.74 (due to the poor positional accuracy of the J1851-0035 calibrator); and $\leq30$ mas for the CRSs associated with G60.88+0.13 and G61.48+0.09.} \\
$^b${Obtained from the 1.3 cm maps of Masqu\'e et al. (2017) convolved to the 7 mm beam size of the maps shown in Figs. \ref{general_maps} and  \ref{detail_maps} (contour maps).}\\
$^c${Spectral index derived using integrated flux.}\\
$^d${Spectral index derived using peak intensity.}
\end{table}


%% file: sources_deconv.tex
\begin{table}[ht]
\footnotesize
\begin{center}
\caption{Physical parameters of the slightly resolved CRSs} \label{deconv}
\begin{tabular}{p{1.7cm}ccccc}
\hline
\hline
               &  $\theta_M \times \theta_m$ ;  P.A $^\mathrm{a}$        &            Size              &  $T_\mathrm{B}^{\mathrm{b}}$ & & $EM^{\mathrm{b}}$     \\
Source   &   ($mas  \times  mas$  ;   $^\circ$)                               &   (AU $\times$ AU)    &       (K) & $\tau^{\mathrm{b}}$  & (10$^8$ cm$^{-6}$ pc)       \\
\hline
G28-VLA1 & $ 114 \pm 15 \times 70 \pm 30   ; 48    \pm $ 20 & $ 340 \pm 50 \times 200 \pm 80$   &  $360\pm 160$   &  $0.036 \pm 0.016$   &  $3.0 \pm 1.3$  \\
G35-VLA1 & $ 61 \pm 8 \times  37 \pm 9        ;136   \pm $ 20 & $ 200 \pm 30 \times 120 \pm 30 $ &  $2300 \pm 600$   &  $0.26 \pm 0.08$   &  $22 \pm 7$\\
G61-VLA1 & $ 155 \pm 23 \times  60 \pm 30  ; 160  \pm $ 8 & $ 310 \pm 50 \times 120 \pm 50 $  &  $240 \pm 120$   &  --  &  --  \\
\hline
\end{tabular}
\end{center}
 $^\mathrm{a}${Size is deconvolved from the beam of each map (see column 4 of Table \ref{maps})} \\
$^{\mathrm{b}}${Brightness temperature (fourth column), optical depth (fifth column) and emission measure (sixth column) obtained in the same fashion as MRT2107, but with the expressions adapted for the frequency of 43.94 GHz (i.e. the central frequency of the observed 7 mm band), and assuming an electronic temperature of 10$^4$ K. }\\

\end{table}